# Strain assisted magnetization switching in ordered nanomagnets of $CoFe_2O_4/SrRuO_3/(Pb(Mg_{1/3}Nb_{2/3})O_3–PbTiO_3)$ hetrostructures

*Anju Ahlawat[1], Azam Ali khan[2,3], Pratik Deshmukh[2,3], Sushmita Bhartiya[4], S. Satapathy[2,3], Mandar.M.Shirolkar[5,6], Haiqian Wang[6], R. J. Choudhary[1]*

[1]UGC-DAE Consortium for Scientific Research, Indore

[2]Laser Biomedical Applications Division, Raja Ramanna Centre for Advanced Technology, Indore, 452013, India

[3]HomiBhabha National Institute, Training School Complex, Anushakti Nagar, Mumbai-400094, India

[4]Laser and Functional Materials Division, Raja Ramanna Centre for Advanced Technology, Indore 452013, India

[5]Symbiosis Center for Nanoscience and Nanotechnology (SCNN), Symbiosis International (Deemed University) (SIU), Lavale, Pune 412115, Maharashtra, India

[6]Hefei National Laboratory for Physical Sciences at the Microscale, University of Science and Technology of China, Hefei, Anhui 230026, China

Corresponding Author: anjahlawat@gmail.com

## Abstract

We have explored the electric field controlled magnetization in the nanodot $CoFe_2O_4/SrRuO_3$/PMN-PT heterostructures. Ordered ferromagnetic CFO nanodots (~300 nm lateral dimension) are developed on the PMN-PT substrate (ferroelectric as well as piezoelectric) using a nanostencil-mask pattering method during pulsed laser deposition. The nanostructures reveal electric field induced magnetization reversal in the single domain CFO nanodots through transfer of piezostrains from the piezoelectric PMN-PT substrate to the CFO. Further, electric field modulated spin structure of CFO nanomagnets is analysed by using X-ray magnetic circular dichroism (XMCD). The XMCD analysis reveals cations ($Fe^{3+}/Co^{2+}$) redistribution on the octahedral and tetrahedral site in the electric field poled $CoFe_2O_4$ nanodots, establishing the strain induced magneto-electric coupling effects. The $CoFe_2O_4/SrRuO_3$/PMN-PT nanodots structure demonstrate multilevel switching of ME coupling coefficient (α) by applying selective positive and negative electric fields in a non-volatile manner. The retention of two stable states of α is illustrated for ~$10^6$ seconds, which can be employed to store the digital data in non-volatile memory devices. Thus the voltage controlled magnetization in the nanodot structures leads a path towards the invention of energy efficient high-density memory devices.

# 1. Introduction:

Currently, a primary goal of the research in materials science is to discover the pathway towards the development of the energy efficient and high-density information storage devices. Utilizing an electric field rather than the usual magnetic field to manipulate the magnetization, provides an efficient method to reduce the power consumption in spintronic devices[1-3]. In this regard, the multiferroic integrated nanostructures (using ferroelectric (FE) and ferromagnetic (FM) materials) which can produce high magneto-electric (ME) coupling effects at room temperature, are appealing [4-8]. Among the various interesting interfacial coupling mechanisms, the elastic coupling mechanism is simple and promising to produce strong ME coupling between two ferroic materials via their strain transfer efficiency across the interfaces [9-12].

Recently, efforts are being made to develop the ordered nano-islands of FE/FM heterostructures to get a sizable ME effect, which can hold tremendous potential in the electric field tuning of magnetic domain switching to achieve the ultimate goal of high-density energy efficient information storage devices[13-15]. A significant ME effect can be produced in the island heterostructures at the nanoscale because of the reduced clamping effect of the substrate, which otherwise deteriorates the ME coupling in multiferroic layered composite films[15-19]. In this regard, various patterning methods such as nanoimprint, anodic aluminum oxide patterns, and nanostencils mask are being explored [20-27].

In such structures, the ME effect can be directly measured as ME coupling coefficient (α)

$$= \left(\frac{dP}{d\lambda_P}\right) * \left(\frac{d\lambda_M}{dH}\right) = dP/dH \qquad \ldots\ldots\ldots\ldots\ldots(1)$$

where $dP/d\lambda_P$ and $d\lambda_M/dH$ where P is electric polarization and $d\lambda_P$ is the change in induced strain via an electric field in the piezoelectric component, H is the applied magnetic field and $d\lambda_M$ refers to the change in a strain which is elastically transferred to the magnetostrictive component (considering $d\lambda_P = d\lambda_M$, i.e. considering perfect elastic strain transfer). The values of α can be utilized to encode the digital information in a non-volatile manner, rather than using the M, P, and resistance (R). The α based non-volatile memories (NVM) propose a few advantages over the existing non-volatile random access memory (RAM)[28-29]. For instance, destructive readout process in FE RAMs can be avoided, where each readout process needs a polarization (P) reversal and then P has to switch back to rewrite the data. However, in case of proposed α based NVMs, no need to change the direction of P and thus the readout process is possible in a non-destructive manner[28-29].

Previously we showed the magneto-electric coupling in the $NiFe_2O_4$/$SrRuO_3$/PMN-PT thin film based systems[13]. It was also shown that the poling caused redistribution of the cations, leading to changes in the magnetic properties as confirmed from the XMCD studies. From the device architecture point of view, we have shown that the nano-dot structures of $NiFe_2O_4$ revealed magneto-electric coupling effect. It should be noted here that magneto-striction is a crucial parameter in determining the magneto-electric properties of strain coupled multiferroic composites. Thus the materials such as $CoFe_2O_4$ possessing higher magneto-striction should be a favourable choice for such studies. Although $NiFe_2O_4$ and $CoFe_2O_4$ exhibit nearly similar crystal structures (Inverse spinel), their magnetic properties are quite different; $NiFe_2O_4$ is a soft ferrimagnet whereas $CoFe_2O_4$ is a hard ferrimagnet material.

In this work, we design ordered nanoislands ME heterostructure consisting of a single-domain arrays of $CoFe_2O_4$ (CFO) nanomagnets fabricated on a ferroelectric substrate. We explore the electric field controlled magnetization reversal within the single-domain CFO nanomagnets. We used a $Si_3N_4$ membrane mask (nanostencils) during pulsed laser deposition (PLD) to develop the periodic array of CFO nanostructures. The 011 $[Pb(Mg_{1/3}Nb_{2/3})O_3]_{0.70}$–$[PbTiO_3]_{0.30}$ (PMN-PT) is chosen as a substrate because it is a robust ferroelectric material with large piezoelectric coefficients[30]. For the 011 $[Pb(Mg_{1/3}Nb_{2/3})O_3]_{0.70}$–$[PbTiO_3]_{0.30}$ (PMN-PT), the piezoelectric coefficients can reach ~ −3100 pC/N ($d_{31}$) along the [100] direction and ~ 1400 pC/N ($d_{32}$) along the [01-1] direction, respectively[30]. The $CoFe_2O_4$ is a primary ferrimagnetic material with huge magnetostriction and a high Néel temperature 840 K [31-32]. $SrRuO_3$ (SRO) is used as top electrode on the PMN-PT substrate. The strain assisted ME effect has been extensively studied via strain transfer from the piezoelectric substrate to the ferromagnetic component, which has made significant progress in the multiferroic heterostructures [33]. Although, a few reports are available on the magneto-electric properties of $CoFe_2O_4$/PMN-PT continuous films[31,32,34,37], however, it is interesting to realize a potentially strong strain-induced ME coupling in the well-ordered CFO/SRO/PMN-PT nanoislands structure, which will have huge implications on designing the energy efficient and high-density memory devices.

## 2. Experimental details:

Ordered $CoFe_2O_4$/$SrRuO_3$/PMN-PT multiferroic nanodot structures were grown by using nanostencil assisted pulsed laser deposition (PLD) technique. A nanostencil mask consist of amorphous silicon nitride ($Si_3N_4$) membranes (200 nm thick) were produced on silicon (Si) wafers, with ordered circular apertures of ~300 nm diameter. The $SrRuO_3$ electrode layer (~ 40 nm) was deposited on (011)-oriented PMN-PT substrate using a KrF excimer laser ($\lambda$ = 248 nm, pulse duration - 20 ns), laser pulse repetition rate of 5 Hz, laser energy density 2.5 J/cm$^2$, target to substrate distance ~ 5 cm, an oxygen partial pressure of 100 mTorr and substrate temperature of 600°C, and. Afterwards, a stencil mask was mechanically attached to $SrRuO_3$/PMN-PT and $CoFe_2O_4$ dots (through nanostencils) were grown at the substrate temperature of 650°C, target to substrate distance ~ 4 cm, laser energy density 2 J/cm$^2$, laser repetition rate of 2 Hz, at oxygen partial pressure ~10-5 Torr1-2. Subsequently, the stencils were unclamped. The schematic of stepwise fabrication process of the films by PLD is shown in as shown in the Fig. 1(a). The $SrRuO_3$ was used as top electrode and gold (Au) layer of thickness ~ 50 nm is deposited on the back side of PMN-PT substrate to serve as bottom electrode for electric field (E) poling. The electric field E= +10 kV/cm (which is above the coercive field of PMN-PT) was applied on the PMN-PT substrate across the thickness.

**Characterization:** The crystalline structure of the nanodots was analyzed by X-ray diffraction using CuK$\alpha$ radiation with wavelength ($\lambda$) =1.54 Å (Bragg–Brentano $\theta$-2$\theta$ geometry). The surface morphology and depth profile of the $CoFe_2O_4$ dots were determined using Asylum atomic force microscope (AFM) equipped Magnetic Force Microscopy (MFM), using low-moment ASYMFMLM Asylum Research tips (25 nm radius) of stiffness 2 Nm$^{-1}$ with CoCr coating (magnetized in a direction perpendicular to the surface of sample). AFM was carried out in the tapping mode and MFM images were recorded in the two-pass mode (to overcome the electrostatic force effect) with lift height of ~50 nm. The calibration of frequency shift associated with the magnetic forces during MFM was carried out using a floppy disk (reference sample).

The magnetic properties of the $CoFe_2O_4$/$SrRuO_3$/PMN-PT nanodot structures were analyzed by superconducting quantum interference device (SQUID) magnetometer. The substrate contribution was removed by subtracting the magnetization of the substrate from the film's magnetization signal. The spin structure of $CoFe_2O_4$ nanodots was examined by element specific X-ray magnetic circular dichroism (XMCD) technique which is carried out in total electron yield mode at polarized light soft x-ray absorption beamline BL-1 at Indus-2 synchrotron source, RRCAT, Indore, India. The XMCD spectra were measured by fixing the incident beam polarization and the magnetic field direction was switched using ± 1 Tesla magnetic field.

Magneto-electric measurements were performed by changing the bias magnetic field ($H_{dc}$) under superimposed in-plane ac magnetic field of 10 Oe (induced by Helmholtz coils), at frequency of 1 kHz. The output voltage was measured using Lock-in amplifier (Stanford, SR532). The reference signal was taken from the signal generator feeding the Helmholtz coils.

## 3. Results and discussion:

Figs 1(a-b) present a schematic of fabrication process by mechanically lifting off the mask, through which well-ordered CFO nanodot array is obtained. Figure 1(c) shows atomic force microscopy (AFM) topography of the CFO nanodots of lateral dimension ~ 300 nm. The AFM depth profile of the CFO/SRO/PMN-PT nanostructures indicates the dot height of ~ 40 nm (Fig. 1(d)). The growth orientation of the CFO nanodots is analyzed by XRD (Fig. 1(e)). The XRD pattern shows only (022), (044) diffraction peaks corresponding to the CFO nanodots and the (011), (022) diffraction peaks corresponding to SRO and the PMN-PT substrate, indicating that the CFO nanodots are oriented along the (011) direction. The inset of Fig. 1(e) shows distinct planes of PMN-PT.

The PMN-PT (011) substrate reveals the remnant polarization of ~25 $\mu C/cm^2$ and coercivity of ≤ 5 kV/cm, respectively[32]. Figure 2 (a-b) shows topography and MFM image of CFO nanodots. The low-moment tips were used to avoid the intervention of the magnetic tip with magnetization of the CFO nanodots. The MFM tip was magnetized perpendicular to the substrate, which is suitable to detect the modifications in the vertical components of magnetic field. Bright and dark contrast in the MFM images of CFO dots indicates repulsion and attraction of the tip to the sample, respectively. The MFM images reveal that the CFO nanodots are magnetized in one direction (designated by arrows in Fig. 2(b)) and exhibit single domain features with distinct North-South poles, as observed in the previous magnetic nanodots also [18]. Figure 2(c) presents magnetization vs magnetic field (M-H) loops (substrate contribution is subtracted) measured in the in-plane direction. It is worth notable that the coercivity ($H_c$) is very high ~ 4 kOe and remanent magnetization is approaching to nearly saturation magnetization, indicating single domain ferromagnetic characteristics of the CFO nanodots at room temperature. Usually the coercivity of CFO is high and it is further very sensitive to the strain and domain configuration in the thin films[34,35].

The electric field induced modifications in the magnetization of the CFO nanodots are analyzed by MFM measurements which are performed before and after applying electric field (E) to the sample. Figure 3(a) reveals schematic illustration of electric poling of $CoFe_2O_4$/$SrRuO_3$/PMN-PT nanodot structures. The topography and magnetic domain images of the CFO nanodots are recorded before applying E (Fig. 3(b-c)). Afterwards, the PMN-PT is electrically poled by applying the electric field (E) = +10 KV/cm. Subsequently, the sample position is relocated (as it was before poling) using topographic features as the reference and magnetic domain imaging is carried out (Fig. 3(c)). After poling, an apparent change is observed in the MFM contrast of CFO nanodots, implying that the orientations of a few domains are switched, which can be visualized by change in contrast of the dots. It can be observed that before applying E, the white-black contrast of the domain of a dot changes to nearly complete black contrast after the electric poling (Fig. (3(b-c)). We observe such domain switching in larger region of the CFO nanodot arrays, however, here we present only a few dots to explicitly reveal switching of the magnetic domains. We shall like to mention here that not all the dot's domain switch under the biasing owing to the presence some intrinsic defects in the films or non-uniform distribution of the induced strain, which may not be sufficiently enough to take over the magnetoelastic energy of the dots.

As PMN-PT is ferroelectric/piezoelectric substrate, the application of electric field generates in-plane anisotropic strain in the PMN-PT, which is then transferred to the CFO nanomagnets through the SRO layer. It has been studied that the strain can be effectively transferred from PMN-PT to magnetic layer through the intermediate electrode layer [13, 36-39]. Positive electric field poling in the PMN-PT induces tension along the [01-1] direction and compression along the [100] direction [39] (the corresponding planes of PMN-PT are shown in the inset of Fig. 1(e)). Since CFO possesses negative magnetostriction, the nanomagnet's magnetization rotates from the initial orientation because when the CFO undergoes tensile strain, the magnetic moments can easily move away from the [01-1] direction [32,40]. Thus after application of voltage, most of the nanodot's magnetization aligns along the direction perpendicular to the plane of film which is reflected from the black spherical dots.

Further, to explore the effect of electric field on the spin structure of CFO nanodots, the film was electrically poled at E= +10 KV/cm and carried out XMCD experiments. Figures 4(a) and 4(b) reveal X-ray absorption and X-ray magnetic circular dichroism (XMCD) spectra measured at applied magnetic field of ±1 Tesla (σ+ and σ−) for Fe $L_{3,2}$ edges of CFO nanodots before and after electric field poling. The XMCD signal reveals typical features corresponding to inverse spinel structure of CFO, which have octahedral [$O_h$] and tetrahedral [$T_d$] sites occupying $Co^{2+}$ and $Fe^{3+}$ cations according to the formula $(Fe^{3+})T_d$ [$Co^{2+} Fe^{3+}$]$O_hO_4$[41]. The positive (P1) and negative (N1 and P2) XMCD features for Fe $L_3$ edge arise due to negative exchange interaction among Fe ions situated at $T_d$ sites ($Fe^{3+}$) and $O_h$ sites ($Fe^{2+}/Fe^{3+}$) respectively.

Further, the cations ($Fe^{+3}$) distribution at the octahedral and tetrahedral sites is estimated using CTM4XAS code[42] (where a linear combination of the simulated XMCD spectra is fitted for the $Fe^{2+/3+}$cations at octahedral and tetrahedral sites). For the un-poled sample, intensity ratio of P1/P2 is 46:54 suggesting that 46% of $Fe^{3+}$ ions reside at $T_d$ sites, and 54 % at the $O_h$ sites, which is nearly close to that expected for the inverse spinel $CoFe_2O_4$[35]. While for the poled samples, intensity ratio of P1/P2 is 21:79, suggesting cations ($Fe^{3+}$) redistribution occurs at $O_h$ and $T_d$ sites. Figure 4(c) clearly reveals that the relative intensities of the P1 and P2 peaks are significantly different in XMCD spectra of Fe $L_{3,2}$ edges of CFO nanodots before and after electric field poling, infering that the occupancy of $Fe^{+3}/Fe^{2+}$ cations on the octahedral ($O_h$) and tetrahedral ($T_d$) sites in the $CoFe_2O_4$ nanodots is modified after electric field poling owing to reduction of a fraction of $Fe^{3+}$ ions into $Fe^{2+}$ ions at $O_h$ site.

Concurrent changes are observed in the XMCD spectra of Co $L_{3,2}$ edges (Fig. 4(d-e)). The intensity of C1 peak is reduced at $O_h$ sites after electric field poling (Fig. 4(f)), suggesting ($Co^{2+}\backslash Fe^{3+}$) cations redistribution at the $O_h$ and $T_d$ sites in the inverse spinel CFO. Thus it can be concluded that the electric field poling causes redistribution of the cations and hence modified the spin structure of the CFO nanodots.

It is known that the application of electric field on the ferroelectric material induces piezoresponse and thus stress builds up on the adjoining layer, which leads to change the magnetic interactions/anisotropy via inverse-magnetostriction [40]. Moreover, polarization of the ferroelectric material give an extra field-effect [34,40]. The electric field driven charge modulation plays an essential role in tuning the magnetism for spinels ferrites via redistribution of cations at $T_d$ or $O_h$ sites. The electric field induced cations ($Fe^{2+\backslash 3+}$, $Co^{2+\backslash 3+}$) redistribution at A and B sites which affects the super exchange interactions (A-O-A, B-O-B and A-O-B) in spinel ferrites which is ultimately correlated with the strength of the exchange interaction among the Fe and Co moments as well as the magnetism [32,40,43]. These modulations will also govern the domains reorientation in the poled CFO nanodots. Hence, the modified magnetism after electric poling of CFO/SRO/PMN-PT nanostructures is due to the strain and field-effect.

Our above results have demonstrated that the domains orientation and the spin structure of the CFO nanomagnets can be significantly modified by electric poling through the converse ME effect [44] in CFO/SRO/PMN-PT nanostructures at room temperature.

Quantitatively, the ME coupling is evaluated by measuring ME coupling coefficient (α) as a function of $H_{dc}$ (in-plane) under the superposition of ac magnetic field ($H_{ac}$) for the CFO/SRO/PMN-PT nanodot structures, as shown in Fig. 5 (a). Before measuring α, the sample was electrically poled by applying positive electric field (+E) to set the polarization directions (↑P) . Then the induced ME voltage was measured as a function of $H_{dc}$ at constant AC magnetic field amplitude and frequency of 1 kHz. The induced ME voltage (α) was measured as $H_{dc}$ scans from +ve to –ve values (black curve). The arrows indicate forward and backward sweeping measurement directions. Then sample was electrically poled with negative positive electric field to reverse the direction of polarization (↓P) and the induced ME voltage (α) was again measured as $H_{dc}$ scans from +ve to –ve values (red curve). The value of α increases with $H_{dc}$ and reaches a peak at $H_{dc} \sim 10$ KOe, where the magnetostriction coefficient of CFO is maximum (because magnetization of CFO is saturated). Then the value of α decreases at high $H_{dc}$ beyond 10 KOe because the peizomagnetic coefficient approaches to nearly zero. Figure 5(a) reveals hysteretic behaviour of the CFO nanodots and non zero α values at $H_{dc}=0$. The maximum values of ME coupling coefficient α = 198 mV/cm-Oe and α = 79 mV/cm-Oe are observed at $H_{dc} \sim 10$ kOe and $H_{dc} = 0$, respectively. The sign of α depends mainly on the relative orientation between P and M, owing to the strain assisted ME coupling across the interfaces [28,29]. For a fixed direction of P, the sign of α can be switched by reversing M with $H_{dc,}$ while for a fixed direction of M, the sign of α can be switched by reversing P with E. For instance, α is > 0 for upward direction of P (+E) and α < 0 for reverse direction of P (-E). Hence the two steady states of α (α > 0 and α < 0) can be achieved by applying ± E, which can be employed to store the digital data in terms of "0" and "1" for non-volatile memory applications, similar to that in FeRAMs[28,29]. The retention of two stable states of α (α > 0 and α < 0) was measured as a function of time. Figure 5(b) demonstrates retention of α for $\sim 10^6$ seconds, measured at $H_{dc} = 10$ kOe (at which maximum values of α is achieved).

In the following, CFO/SRO/PMN-PT nanodot structures demonstrate electric field controlled multilevel switching of α with respect to (w.r.t) alternate applied electric pulses (Fig. 5(c)). As the ferroelectric are known to posses intermediate P states between two saturation values (± Ps) by altering the ratio of up and down ferroelectric domains [45]. Consequently, the ME coefficient α can also has different values for a fixed direction of M. For the first cycle, an electric pulse of +5 kV/cm is applied (+P) and α was measured for 50s; after that E= -5 kV/cm was applied and α was measured for another 50s. Then the measurements are performed for many cycles with the different positive values of E (+ 4.5 kV/cm, +4 kV/cm and +3.5 kV/cm); whereas negative pulses of E = - 5 kV/cm is applied for each cycle. When the applied E reverses P, α reverses its sign and retains its state until the next electric pulse is applied. We observe different levels of α by applying distinct +E pulse owing to either fully or partially reversal of the ferroelectric domains. After each *E* pulse, α retains its state without apparent decay. The multilevel switching measurements are performed at dc bias zero ($H_{dc}$=0) as it is beneficial to operate without a dc bias for practical applications. The CFO/SRO/PMN-PT nanodot structures illustrate steady switching features for future NVM applications.

It is worth noticeable that the non-volatile change of magnetoelectric coupling coefficient α is not caused by the memory effect of the electric field induced strain. In the ferroelectric/ferromagnetic composite systems, the M and P are coupled via strain and induced ME voltage α is a consequence of the coupling of two polarization states (M and P). Hence in such ferroelectric/ferromagnetic composite systems, the memory effect is not built by the single polarization process.

The observed results in ordered nanomagnet structures will have huge implications on the read out process in the magnetic RAMs, which is comparatively simple and non destructive as compared to ferroelectric RAMs. In the proposed α based memories, the read out process can be achieved by measuring the sign of the induced ME voltage (ΔV). During this process a small ac magnetic field (~ 1 Oe) is applied which is produced by an independent coil (as shown schematically in Fig (5(d)). Hence, all the stored digital data can be read out in parallel leading to the enhanced data density and lower power consumption.

## 4. Conclusion:

$CoFe_2O_4$/SRO/PMN-PT nanodot structures are developed by a stencil-based patterning method and the ME coupling effect is explored at nanoscale. AFM measurements reveal the ordered growth of CFO nanodots. The electric field controlled non-volatile modification in the magnetism of CFO nanodots array is observed in the $CoFe_2O_4$/$SrRuO_3$/PMN-PT heterostructures. MFM reveals electric field induced change in the domain orientation of CFO nanodots. XMCD studies confirm modulation in the spin structure of CFO nanodots before after the electric field poling. The electric field controlled multilevel switching of α in the ordered ME nanodots structures offers exciting opportunities towards energy efficient non-volatile memories.

**Acknowledgement:** Authors acknowledge Mr. A. Wadikar and Mr. R. K. Sah and for their help in the X-ray absorption measurements. Authors are grateful to CSIR, New Delhi for providing financial assistance.

**Data Availability Statement:** The data which supports findings of the present study are available within the article.

**Figure captions:**

**Figure 1**. Schematic of $CoFe_2O_4$ nanodots patterning on $SrRuO_3$-coated single crystal PMN-PT (011) (a) deposition of $CoFe_2O_4$ through nanostencil-mask, (b) lifting of the mask (c-d) Atomic force microscopy image and measured height profile along with the white line (shown in inset) for the ordered circular $CoFe_2O_4$ nanodots, respectively, (e) XRD pattern of $CoFe_2O_4$ nanodots/$SrRuO_3$/PMN-PT heterostructures, the inset shows distinct planes of PMN-PT.

**Figure 2**. (a) Topographic features (b) MFM image of $CoFe_2O_4$ nanodots, arrows point towards orientation of magnetic domains (c) M-H loops measured for $CoFe_2O_4$/$SrRuO_3$/PMN-PT heterostructures.

**Figure 3.** (a) Schematic demonstration of electric field poling of the $CoFe_2O_4$/$SrRuO_3$/PMN-PT nanodots structures, (b) AFM topography (c-d) corresponding MFM image of the $CoFe_2O_4$ dots before and after electric field poling, respectively, arrows indicate change in magnetic domain orientation of $CoFe_2O_4$ dots.

**Figure 4:** XAS and XMCD measured for $L_{3,2}$ edges (a) before electric poling (b) after electric poling, (d-e) $L_{3,2}$ edges for co edge of $CoFe_2O_4$ dots before and after electric field poling, respectively, (c-f) comparative $L_3$ edges XMCD signal for Fe and co edge of $CoFe_2O_4$ dots before and after electric field poling, respectively.

**Figure: 5** (a) Schematic configuration of the ME coefficient measurement and the measured ME coupling coefficient (α) as a function of $H_{dc}$ for $CoFe_2O_4$/$SrRuO_3$/PMN-PT samples, (b) time dependent retention of α, (c) multilevel switching of α as a function of time by applying ±electric pulses (E) at $H_{dc}= 0$, (d) illustration of reading operation, the ordered nanodot structures are put into a reading coil which induce a small magnetic field ΔH. The stored data (in terms of α) can be read out by recording the sign of induced ME voltage (ΔV).

**Figure 1**

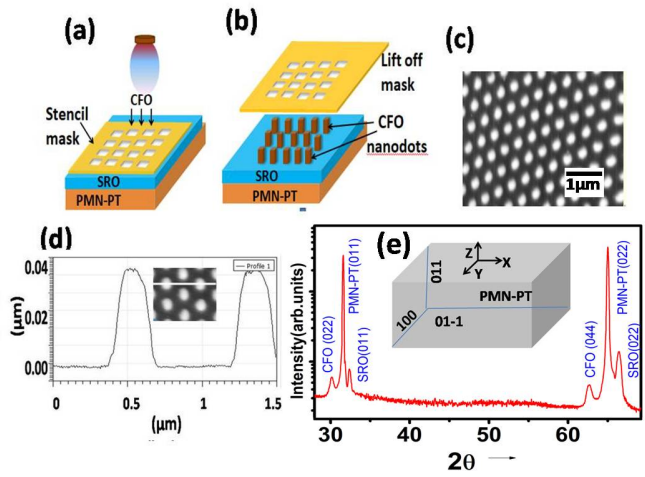

**Figure 2**

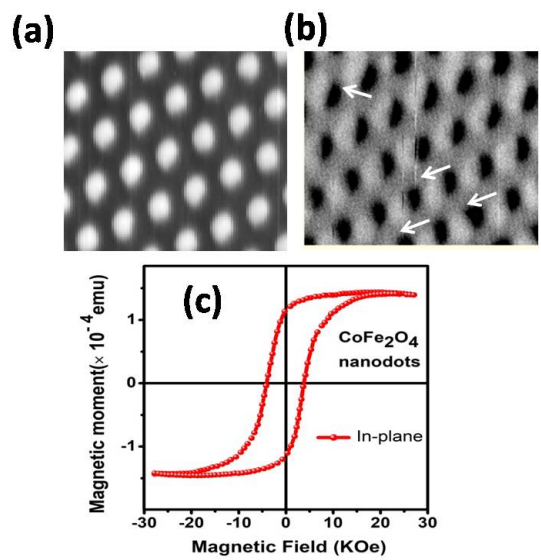

**Figure 3**

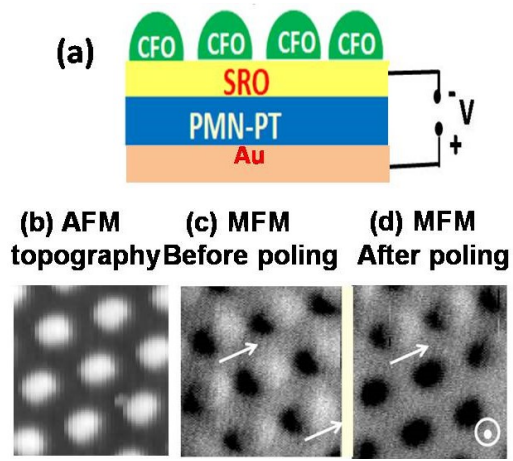

**Figure 4**

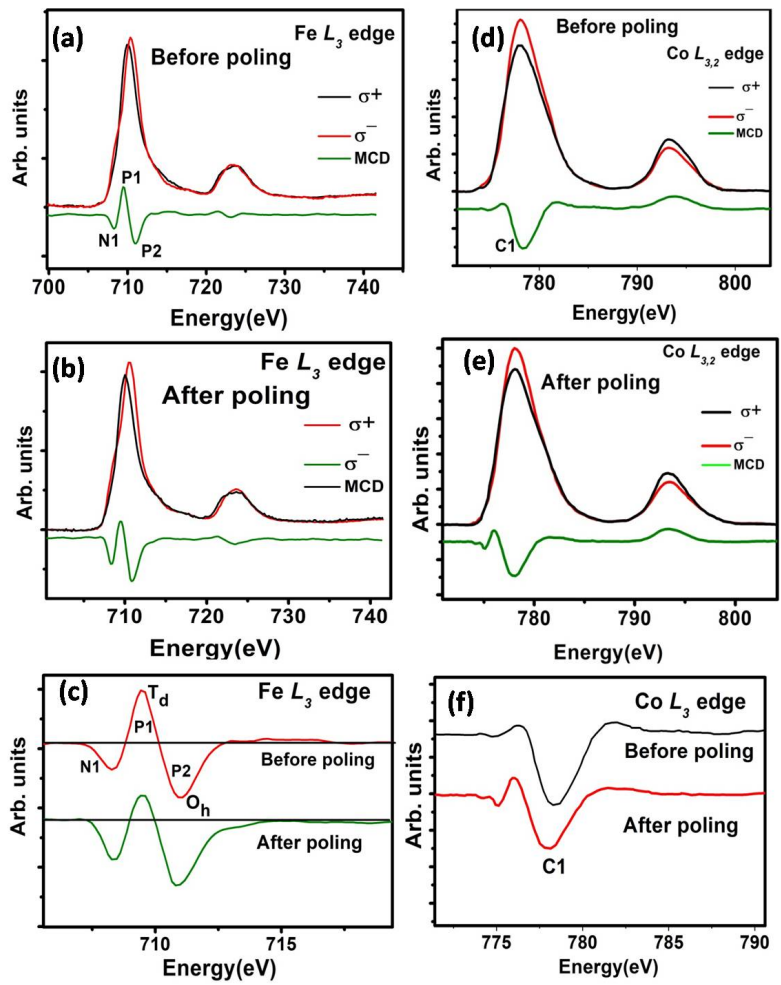

**Figure 5**

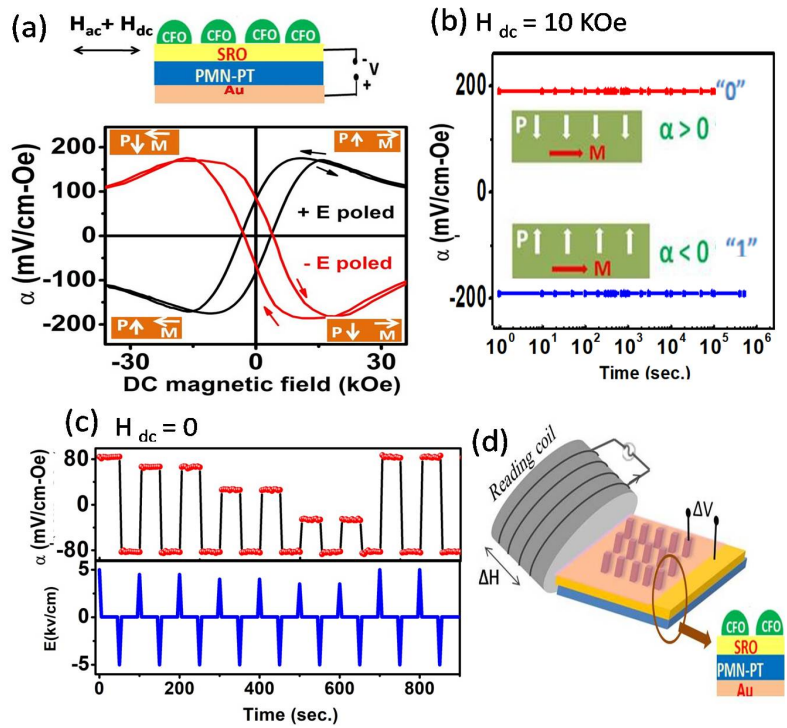